\documentclass[fleqn,10pt]{wlscirep}
\usepackage[utf8]{inputenc}
\usepackage[T1]{fontenc}
\usepackage{hyperref}

\title{Commuting network effect on urban wealth scaling}
\author[1]{Luiz G. A. Alves}
\author[2,3,4]{Diego Rybski}
\author[5,*]{Haroldo V. Ribeiro}

\affil[1]{Department of Chemical and Biological Engineering, Northwestern University, Evanston, IL 60208, U.S.A.}
\affil[2]{Potsdam Institute for Climate Impact Research -- PIK, Member of Leibniz Association, P.O.\ Box 601203, 14412 Potsdam, Germany}
\affil[3]{Institute for Environmental Science and Geography, University of Potsdam, 14476 Potsdam, Germany}
\affil[4]{Department of Environmental Science Policy and Management, University of California Berkeley, 130 Mulford Hall \#3114, Berkeley, CA 94720, USA}
\affil[5]{Departamento de F\'isica, Universidade Estadual de Maring\'a, Maring\'a, PR 87020-900, Brazil}

\affil[*]{Correspondence to hvr@dfi.uem.br}

\begin{abstract}
Urban scaling theory explains the increasing returns to scale of urban wealth indicators by the per capita increase of human interactions within cities. This explanation implicitly assumes urban areas as isolated entities and ignores their interactions. Here we investigate the effects of commuting networks on the gross domestic product (GDP) of urban areas in the US and Brazil. We describe the urban GDP as the output of a production process where population, incoming commuters, and interactions between these quantities are the input variables. This approach significantly refines the description of urban GDP and shows that incoming commuters contribute to wealth creation in urban areas. Our research indicates that changes in urban GDP related to proportionate changes in population and incoming commuters depend on the initial values of these quantities, such that increasing returns to scale are only possible when the product between population and incoming commuters exceeds a well-defined threshold.
\end{abstract}
\begin{document}

\flushbottom
\maketitle

\thispagestyle{empty}

\section*{Introduction}
Cities are engines of innovation~\cite{hong2020universal}, happiness~\cite{glaeser2012triumph}, creativity~\cite{bettencourt2007growth}, production of knowledge~\cite{nomaler2014scaling} and wealth~\cite{west2017scale}, but are also associated with the spread of diseases~\cite{ribeiro2020city}, violent crimes~\cite{alves2013distance, alves2018crime, alves2015scale}, segregation~\cite{sampson1997neighborhoods, louf2016patterns}, and pollution~\cite{barthelemy2019statistical, ribeiro2019effects}. The prominent role of urban systems in the society and the increasing number of people living in cities have driven significant and ongoing efforts towards developing a new science of cities~\cite{batty2008size, batty2013new, barthelemy2019statistical, lobo2020urban}, which has been further fostered by the increasing availability of highly detailed city data.

Urban scaling is among the most remarkable and widely acclaimed findings of this new science of cities~\cite{bettencourt2007growth}. This theory states that different urban indicators ($Y$) can be expressed by a power-law function of city population ($N$), that is, $Y\sim N^\beta$, where $\beta$ is the urban scaling exponent or the scale-invariant elasticity. A direct consequence of this nonlinear relationship is that percentage increments in population may result in greater or smaller percentage changes in urban indicators depending on whether $\beta>1$ (increasing returns to scale) or $\beta<1$ (decreasing returns to scale). The latter case is well illustrated by economies of scale in material infrastructure of large cities~\cite{bettencourt2007growth}, while the more than proportional increase of urban wealth with city population is a typical example of the former case~\cite{bettencourt2007growth, sveikauskasproductivity1975}.

In an analogy with the origins of allometric scaling in biology~\cite{west1997general}, the emergence of these urban scaling laws has been attributed to intra-city networks~\cite{bettencourt2013origins}. The underlying infrastructure network of cities (such as electric cable and road networks) plays the role of the biological networks (such as circulatory and respiratory networks), and network constraints would explain the decreasing returns to scale observed in infrastructure indicators as well as in animals metabolic rates. However, this analogy is limited as there is no biological counterpart for the emergence of increasing returns to scale in urban socioeconomic indicators. The per capita increase of social and economic quantities with city population is usually attributed to the superlinear scaling of human interaction networks within cities~\cite{arbesman2009superlinear, pan2013urban, bettencourt2013origins, schlapfer2014scaling}, that is, to the finding that per capita levels of individual contacts and communication activities also increase with population.

Urban scaling theory implicitly assumes cities as isolated entities and that only within-city interactions would be responsible for the increasing returns to scale of city socioeconomic indicators. This assumption contrasts with the complex inter-city interactions~\cite{gonzalez2008understanding, brockmann2006scaling, masucci2013gravity, louf2013modeling, ren2014predicting, spadon2019reconstructing, keuschnigg2019urban, bettencourt2019demography, altmann2020spatial} that manifest through people's everyday movements from home to work (commuting networks) and long-term migration patterns among cities. These networks allow cities to attract people from different neighboring areas, contribute to information flow, city development, and the formation of more creative and productive teams~\cite{guimera2005team}. Despite the importance of inter-city interactions, there have been very few attempts to put forward the more holistic idea that urban scaling may not only emerge from processes occurring within the city boundaries~\cite{ribeiro2021association}.

Here we investigate the effects of commuting networks on the urban gross domestic product (GDP) scaling using data from the US and Brazil. To do so, we use an analogy with the economic theory of production functions~\cite{heathfield1987introduction} proposed by Ribeiro \textit{et al.}~\cite{ribeiro2019effects} to account for the interconnected role of population and incoming commuters on urban wealth creation. Our results show that modeling urban GDP as a function of population and number of incoming commuters significantly improves the description of urban productivity when compared with usual urban scaling. These models explicitly consider interactions between population ($N$) and number of incoming commuters ($S$) and indicate that the elasticity of scale of urban GDP depends on the product of these two quantities ($\Omega = N\times S$). This dependence implies the existence of a transition from decreasing to increasing returns to scale when $\Omega$ exceeds a particular threshold. Our research suggests that cities increase their productivity from a combination of intra- and inter-city processes but that increasing returns to scale of urban GDP only emerge when the intensity of the product of these processes exceeds a threshold. Therefore, the GDP of two cities with similar populations may respond differently to a percentage change in population if their number of incoming commuters is different. The comparison between the results for the US and Brazil further suggests that the elasticity of scale also depends on the development degree of the urban area. 

\section*{Results}

\begin{figure}[!t]
\centerline{\includegraphics[width=.9\columnwidth]{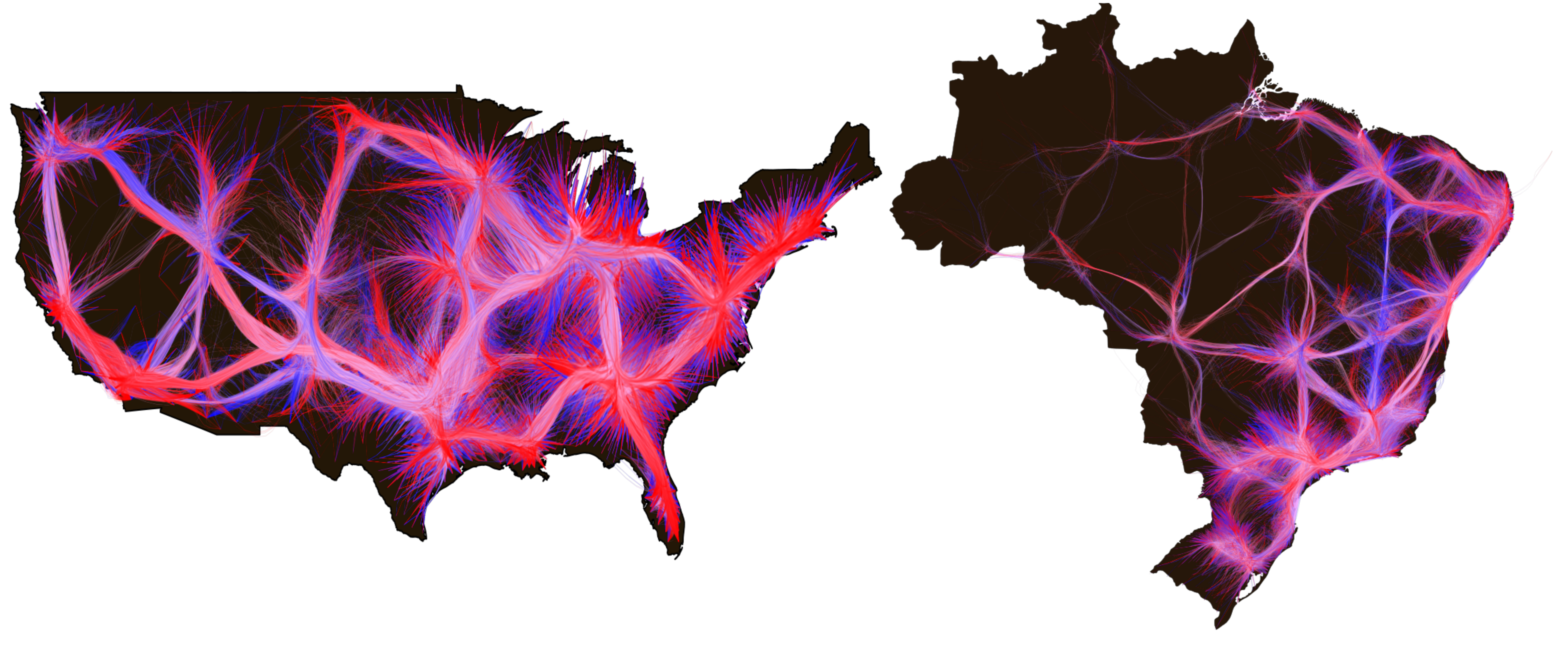}}
\caption{Visualization of the commuting networks among US counties (left) and Brazilian (right) municipalities. In this representation, nodes are urban areas and links represent the flow of commuters among them. Edge widths are associated with the flow intensity and colors refer to flow direction (bluish tones indicate outgoing links while reddish indicate incoming connections). This figure has been made using matplotlib~\cite{hunter2007matplotlib} and geopandas~\cite{jordahl2014geopandas}, we  have further aggregated links based on their spatial proximity using a kernel-based edge bundling algorithm~\cite{moura20153d}. The overall complexity of interactions among urban areas motivates investigations about the role of these networks on urban scaling.}
\label{fig:fig1}
\end{figure}

We start our investigation by visualizing the complex network related to people's everyday movements from home to work. We have collected data from the US Census Bureau and the Brazilian Institute of Geography and Statistics about the commuting flow among urban areas in the US and Brazil (see Methods for details). The urban units comprise 3,220 US counties and 5,565 Brazilian municipalities. Figure~\ref{fig:fig1} shows these spatial networks where vertices are urban areas (US counties or Brazilian municipalities), and links represent the flow of workers among these urban areas. Links are directed from home to workplace, and their weights stand for the total number of workers along a given path. The complexity of these commuting networks contrasts with the current view of urban scaling theory that assumes cities as isolated entities and intra-city processes as the only ones responsible for urban scaling. It seems natural to assume that the flow of people among urban areas contributes, at least to a certain degree, to wealth and knowledge creation in cities; however, this possibility remains little explored~\cite{ribeiro2021association}.

\begin{figure*}[t]
\centerline{\includegraphics[width=1\textwidth]{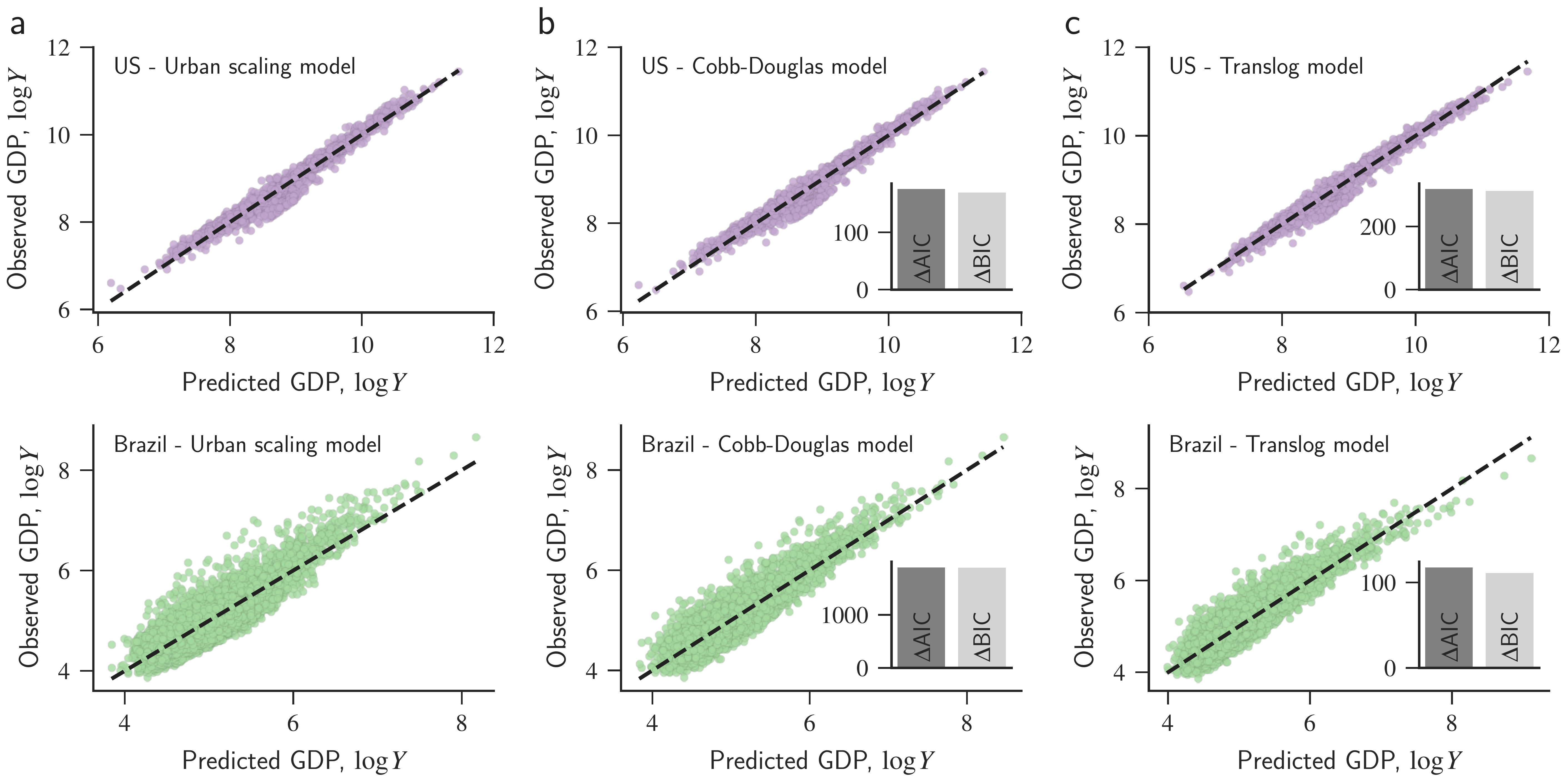}}
\caption{Urban scaling predictions of GDP and the improved descriptions based on the interplay between population and incoming commuters. Relation between the observed values of urban GDP and those predicted by (a) usual urban scaling (Eq.~\ref{eq:urban_scaling}), (b) Cobb-Douglas model (Eq.~\ref{eq:cobb_douglas_scaling}), and (c) translog model (Eq.~\ref{eq:translog_scaling}). All quantities are expressed in base-10 logarithmic scale and the dashed lines represent the 1:1 relationship. The upper plots (purple markers) show the results for US counties, and the lower ones (green markers) represent the values for Brazilian municipalities. The bar plot inserted in panel (b) shows the differences in the Akaike information criterion ($\Delta \text{AIC}$) and Bayesian information criterion ($\Delta \text{BIC}$) estimated between the Cobb-Douglas and urban scaling models. The large positive values of these quantities indicate that the Cobb-Douglas model significantly improves the description of urban GDP when compared with usual urban scaling. The inset in panel (c) shows the same quantities ($\Delta \text{AIC}$ and $\Delta \text{BIC}$) calculated between the translog and the Cobb-Douglas models, where the large positive values indicate that the translog model further refines the goodness of fit.}
\label{fig:fig2}
\end{figure*}

Before we approach this problem, let us first revisit the urban scaling of GDP. To do so, we have also obtained population and GDP data for all US counties and Brazilian municipalities in our data set (see Methods for details). The urban scaling (also known as the Bettencourt-West law)~\cite{bettencourt2007growth, batty2013new} states that GDP ($Y$) and population ($N$) of urban areas are related via a power-law function
\begin{equation}\label{eq:urban_scaling}
    Y \sim N^\beta \quad \text{or} \quad \log Y \sim \beta \log N \,,
\end{equation}
where $\beta$ is the urban scaling exponent. Furthermore, urban scaling theory predicts $\beta$ to be larger than one for GDP and other socioeconomic indicators~\cite{bettencourt2013origins}. This would thus explain the agglomeration economies in cities, that is, the more than proportional increase of urban wealth with city population. We estimate the values of $\beta = 1.04 \pm 0.003$ for US counties and $\beta = 1.04 \pm 0.01$ for Brazilian municipalities via usual least-squares method (see Fig.~S1). These values are statistically larger than one ($p-$values~$=0$, permutation test) and agree with urban scaling theory in the sense that socioeconomic indicators are expected to display a superlinear relation with city population~\cite{bettencourt2007growth, bettencourt2013origins}. These results further indicate that a 1\% increase in the population of an urban area implies an increase of 1.04\% in its GDP, both for the US and Brazil. Figure~\ref{fig:fig2}{a} shows the relationship between the observed values of urban GDP and those predicted by urban scaling (Eq.~\ref{eq:urban_scaling}).

Despite the good quality of the predictions based on Eq.~\ref{eq:urban_scaling}, we observe some systematic deviations as urban scaling tends to underestimate the urban GDP for large values of this quantity (especially for Brazilian municipalities). Furthermore, the universality of these scaling exponents has been debated, and results indicate that their values vary from country to country~\cite{ribeiro2021association}, are affected by spatial distributions of cities~\cite{altmann2020spatial}, and susceptible to different city definitions~\cite{arcaute2015constructing, alvioli2020administrative, alvioli2020comparative}. In this regard, it is worth noticing that the exponent reported here for US counties is smaller than the results obtained when considering the urban units defined in terms of US metropolitan statistical areas~\cite{bettencourt2013origins} and close to the value obtained when defining US cities as functional urban areas~\cite{ribeiro2021association}. The combination of these caveats with the overall complexity of commuting networks thus invites more complex modeling approaches of urban GDP that account for additional dependent variables beyond urban population. In this context, an interesting analogy with the economic theory of production functions~\cite{heathfield1987introduction} has been recently proposed for modeling carbon dioxide emissions while simultaneously considering the effects of population and area~\cite{ribeiro2019effects}. 

Inspired by the approach of Ref.~\cite{ribeiro2019effects}, we propose to model urban GDP ($Y$) as the output of a two-factor production model where population ($N$) and incoming commuters ($S$) in a given urban area are the inputs of this hypothetical production process. Mathematically, we write $Y = F(N, S)$, where $F$ represents the form of the production function that we borrow from the theory of production functions~\cite{heathfield1987introduction}. Just like combinations of capital and labor may lead to the similar output of goods in a factory, our approach thus considers that a city generates wealth through a process that is mediated by its population (intra-city process) and by the population of workers that is attracted from other cities (inter-city process). It is worth noticing that the number of incoming or outgoing commuters are the simplest features we can extract from commuting networks (in network language they represent the in- and out-strength of nodes~\cite{newman2010networks}, respectively). Beyond a simplifying condition, our choice for considering only the number of incoming commuters as an input factor of the production function relates to the fact that this variable can be more directly associated with cities' importance and development degree. We expect a city that attracts many workers from neighboring cities to have a more diverse and developed labor market than a counterpart with few incoming commuters. A city with a large number of incoming commuters is also likely to affect and exert a large influence on its neighborhood, such that this variable can be thought of as an indirect proxy for quantifying the degree of interaction among cities. We further believe the incoming commuters variable fits nicely with the analogy of production functions and allows a direct interpretation of substitution effects between internal and commuting populations on urban wealth production. Moreover, we have verified that the numbers of incoming and outgoing commuters are strongly correlated, and the inclusion of outgoing commuters in the models we shall discuss does not statistically improve the description of urban GDP.

We first consider the Cobb-Douglas model~\cite{cobb1928theory}
\begin{equation}\label{eq:cobb_douglas_scaling}
Y \sim N^{\beta_N} S^{\beta_S } \quad \text{or} \quad \log Y \sim \beta_N \log N + \beta_S \log S\,,
\end{equation}
where $\beta_N$ and $\beta_{S}$ are two independent parameters. This model is one of the most widely used production functions~\cite{heathfield1987introduction} and can be thought of as a generalization of Eq.~\ref{eq:urban_scaling} that is recovered by setting $\beta_N=\beta$ and $\beta_{S}=0$. Similarly to usual urban scaling, the Cobb-Douglas production function has a scale-invariant elasticity of scale equal to $\varepsilon = \beta_N + \beta_S$~\cite{heathfield1987introduction}. This means a proportionate change in population ($N$) and incoming commuters ($S$) associates with a proportionate change in urban GDP independent of $N$ and $S$. Moreover, we have increasing returns to scale when $\beta_N + \beta_S > 1$ (doubling $N$ and $S$ implies more than doubling $Y$), decreasing returns to scale when $\beta_N + \beta_S < 1$ (doubling $N$ and $S$ implies less than doubling $Y$), and constant returns to scale when $\beta_N + \beta_S = 1$ (doubling $N$ and $S$ implies doubling $Y$).

We can interpret Eq.~\ref{eq:cobb_douglas_scaling} as the result of including incoming commuters as a confounding variable in the usual urban scaling (Eq.~\ref{eq:urban_scaling}) via a multiple linear regression in the log-transformed variables. Thus, if the Cobb-Douglas model represents a significantly better description of data, we cannot reject the hypothesis that incoming commuters contribute to the deviations observed in the usual urban scaling and ultimately to the GDP of an urban area. We have fitted the linearized version of Eq.~\ref{eq:cobb_douglas_scaling} to data via the ridge regression approach~\cite{hoerl1970ridge, friedman2001elements}. As detailed in Methods, using this regularization method is essential for accounting for the multicollinearity between the predictor variables ($\log N$ and $\log S$) and yielding stable estimates for the parameters $\beta_N$ and $\beta_S$. This approach yields $\beta_N = 0.93 \pm 0.01$ and $\beta_S = 0.10\pm 0.01$ for US counties, and $\beta_N = 0.75 \pm 0.01$ and $\beta_S = 0.33 \pm 0.01$ for Brazilian municipalities. All parameters are statistically different from zero in both data sets ($p-$values~$<0.05$, permutation test). Figure~\ref{fig:fig2}{b} compares the observed values of urban GDP for US counties and Brazilian municipalities with the predictions of the Cobb-Douglas model. We note that this model improves the description of urban GDP for both countries and reduces the bias for large GDP values observed for usual urban scaling. Furthermore, the Akaike and Bayesian information criteria (insets of Fig.~\ref{fig:fig2}{b}) indicate that it is much more likely that the empirical data come from the Cobb-Douglas model (Eq.~\ref{eq:cobb_douglas_scaling}) than from the usual urban scaling relation (Eq.~\ref{eq:urban_scaling}).

The significantly better description of data by the Cobb-Douglas model and the fact that $\beta_N < \beta$ for both data sets indicates that urban scaling is considerably affected by the confounding factor of incoming commuters. Moreover, we find $\beta_N+\beta_S>1$ for both countries such that every 1\% increase in population and incoming commuters associates with 1.03\% and 1.08\% raises in the GDP of US counties and Brazilian municipalities, respectively. While the precise comparison between the results of both countries demands a unified city definition (which is not our case), the significantly different values of $\beta_S$ suggest that the GDP of urban areas in Brazil is more affected by commuters than in the US case. Indeed, the Cobb-Douglas model predicts that a 1\% increase in incoming commuters (keeping population unaltered) implies a 0.33\% rise in the GDP of Brazilian municipalities, while the GDP of US counties is expected to increase only 0.10\%. This result suggests that the US urban labor market is significantly more integrated than the Brazilian one; that is, the US commuting network is in a much more optimized state such that changes in the number of commuters have a small effect on urban GDP. On the other hand, the developing stage of Brazil is likely to impose considerable infrastructure limitations that hamper its commuting network to achieve a similar optimal state. These limitations are likely to disproportionately affect small urban areas rather than large urban centers, which in turn may explain the more significant dependence of urban GDP on incoming commuters observed in Brazil. Thus, the larger value of $\beta_S$ for Brazil may represent a latent possibility for increasing the overall GDP of Brazilian municipalities (especially the small ones) that seems limited by transportation infrastructure.

Although the Cobb-Douglas model represents a better description of our data, it also imposes some limitations to the productivity of cities (as well as to firms~\cite{heathfield1987introduction}) that may not necessarily hold. One of these limitations is associated with the elasticity of scale ($\varepsilon$) independent of population and incoming commuters. Another one is related to the so-called elasticity of substitution~\cite{heathfield1987introduction}. This quantity represents the ratio between a proportionate change in the inputs and the associated proportional change in the slope of an isoquant of $F(N, S)$, that is, the elasticity of substitution measures the efficiency at which we can substitute population by incoming commuters while keeping the same level of GDP. The elasticity of substitution is independent on $N$, $S$, and $Y$, and always unitary in the Cobb-Douglas model~\cite{heathfield1987introduction}.

To overcome these limitations and introduce more flexibility to our modeling of urban GDP, we have considered the transcendental logarithmic (translog) production function~\cite{christensen1973transcendental, heathfield1987introduction}
\begin{equation}\label{eq:translog_scaling}
    \log Y \sim \beta_N \log N + \beta_S \log S + \beta_C \log N \log S\,,
\end{equation}
where $\beta_C$ is an additional parameter. We observe that the translog model generalizes the Cobb-Douglas function (recovered by setting $\beta_C=0$) by including an interaction term between population and incoming commuters. This interaction allows population effects on GDP to depend on the number of commuters as well as commuters effects on GDP to depend on population. Indeed, a small proportionate increment in population [$N \to (1+x)N$, with $x \ll 1$] implies a proportionate increment in GDP that depends on incoming commuters [$Y \to (1+x(\beta_N + \beta_C \log S))Y$]. Similarly, a small proportionate increment in incoming commuters [$S \to (1+x)S$] also results in a proportionate increment in GDP that depends on population [$Y \to (1+x(\beta_S + \beta_C \log N))Y$]. The translog model has an elasticity of scale $\varepsilon = \beta_N + \beta_S + \beta_C \log(N S)$~\cite{heathfield1987introduction} that depends on population and incoming commuters. Thus, variations in GDP associated with proportionate changes in population and incoming commuters depend not only on the magnitude of these changes but also on the initial values of $N$ and $S$. All these features are not present in usual urban scaling (Eq.~\ref{eq:urban_scaling}) nor in Cobb-Douglas model (Eq.~\ref{eq:cobb_douglas_scaling}), as both models implicitly assume changes in urban GDP to be scale invariant. The translog model relaxes this hypothesis and adds more flexibility to describe urban GDP in terms of population and incoming commuters.

We have fitted the translog model to our data with the ridge regression approach (see Methods for details), finding $\beta_N = 0.76 \pm 0.01$, $\beta_S = -0.14 \pm 0.01$, and $\beta_C = 0.05 \pm 0.002$ for US counties, and $\beta_N = 0.62 \pm 0.01$, $\beta_S = -0.01 \pm 0.03$, and $\beta_C = 0.08 \pm 0.01$ for the Brazilian municipalities. Except for $\beta_S$ in the Brazilian data set, all model coefficients are statistically different from zero ($p-$values~$<0.05$, permutation test). It is worth noting that $\beta_S\approx0$ does not mean incoming commuters have no role on the GDP of Brazilian municipalities, but that the interaction term related to population and incoming commuters is more significant than the isolated effect of commuters. On the other hand, the negative value of $\beta_S$ obtained for the US data indicates that increasing the number of incoming commuters has a negative effect on the GDP of small counties. Indeed, by rewriting Eq.~\ref{eq:translog_scaling} as $\log Y \sim \beta_N \log N + (\beta_S + \beta_C \log N) \log S$, one readily notes that an increase in $S$ associates with a reduction in $Y$ when $N<N^*$ with $N^* = 10^{-\beta_S/\beta_C}$. We have $N^* \approx 631$ for US counties, which in turn allows a negative association between GDP and incoming commuters for tiny urban areas. In practice, this negative association would be limited to the eleven US counties having population smaller than 631 people.

Figure~\ref{fig:fig2}{c} shows the relation between the actual values of urban GDP for the US and Brazil and the predictions based on the translog model. We have verified that the translog model (Eq.~\ref{eq:translog_scaling}) improves the description of urban GDP in both data sets when compared with the usual urban scaling (Eq.~\ref{eq:urban_scaling}) and the Cobb-Douglas production function (Eq.~\ref{eq:cobb_douglas_scaling}), reducing even further the biases observed for usual urban scaling. The Akaike and Bayesian information criteria (insets of Fig.~\ref{fig:fig2}{c}) indicate it is much more likely that the empirical data come from the translog model than from the Cobb-Douglas model; consequently, data is even more likely to come from the translog model than from usual urban scaling.

In addition to representing a significantly better description for the GDP of US counties and Brazilian municipalities, the adjusted translog models indicate that the effects of population and incoming commuters on urban GDP intensifies with the increase of the product $\Omega = N S$ (that is, $\beta_C>0$). The elasticity of scale is $\varepsilon = 0.62 + 0.05 \log(N S)$ for US counties and $\varepsilon = 0.61 + 0.08 \log(N S)$ for Brazilian municipalities. Thus, the greater the population and incoming commuters of a city, the larger the impact of proportionate changes of these quantities on its urban GDP. This effect is also more intense for Brazilian municipalities than US counties, such that urban areas of both countries with similar populations and incoming commuters respond differently to proportionate changes in these quantities, with urban areas in Brazil reporting proportionally more GDP gain than the US ones. It is worth noticing that this behavior is not related to raw values of urban GDP (which are almost always higher in the US than in Brazilian urban areas); instead, elasticity of scaling is associated with proportionate changes in GDP. A similar result was observed in the context of usual urban scaling of GDP for West and East cities of the European Union, where West cities (usually richer and more developed than their East counterparts) display smaller increasing returns to scale than East cities~\cite{strano2016rich}. Again, we can associate this behavior with the developing nature of Brazilian municipalities in contrast with the more steady economic situation of US counties.

\begin{figure*}[!t]
\centerline{\includegraphics[width=0.8\textwidth]{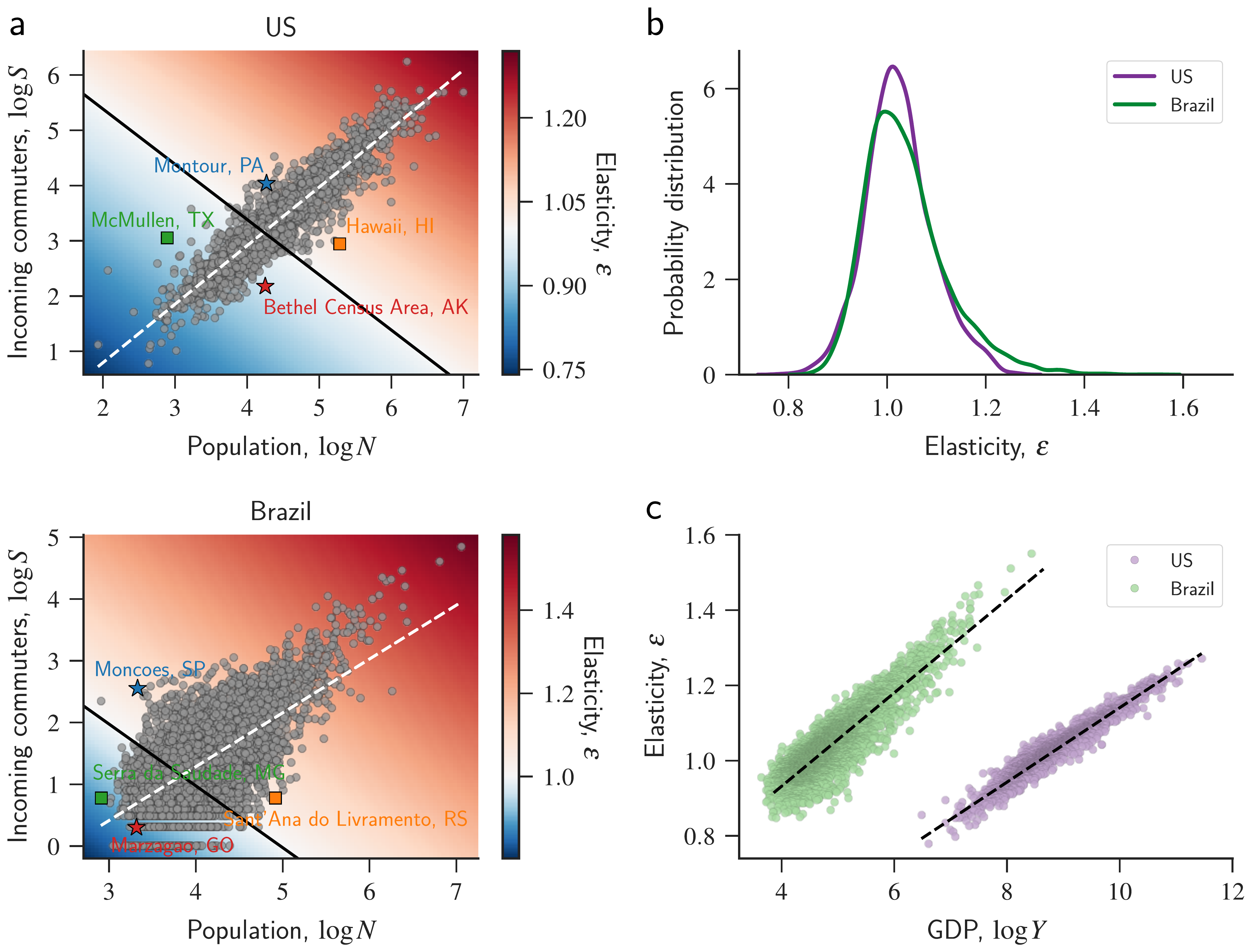}}
\caption{Elasticity of scale and the transition from decreasing to increasing returns to scale of GDP. (a) Dependence of the elasticity of scale ($\varepsilon$) on population ($N$) and incoming commuters ($S$) calculated from the translog model for US counties (upper panel) and Brazilian municipalities (lower panel). The color-coded background represents the elasticity of scale for each location in the plane population versus incoming commuters (on base-10 logarithmic scale) and the gray circles indicate the values of $\log N$ versus $\log S$ for each US and Brazilian urban area. The solid lines in both panels separate the region of decreasing returns to scale ($\varepsilon<1$: bluish region) from the increasing returns to scale one ($\varepsilon>1$: reddish region). We notice that urban areas are spread around the trend defined by $S \sim N^\gamma$ (or $\log S \sim \gamma \log N$, with $\gamma = 0.87 \pm 0.01$ for the US and $\gamma = 1.06 \pm 0.01$ for Brazil -- white dashed lines) and that about 63\% of urban areas in the US and Brazil display increasing returns to scale. The spread of the relationship between $\log S$ and $\log N$ allows urban areas with similar population sizes (as illustrated by the star markers) or similar incoming commuters (as illustrated by the square markers) to have significantly different values of elasticity of scale. (b) Probability distribution of the elasticity of scale estimated for all urban areas in the US (purple) and Brazil (green). The average elasticity of scale $\langle \varepsilon \rangle$ is slightly larger for Brazil than the US ($\langle \varepsilon \rangle=1.04\pm0.001$ and $\langle \varepsilon \rangle=1.02\pm0.001$, respectively), while the Brazilian distribution is significantly more skewed to the right when compared with the US case (skewness~$=1.1$ and skewness~$=0.3$, respectively). (c) Relationship between the elasticity of scale and the GDP of urban areas in the US (purple) and Brazil (green). The GDP of Brazilian municipalities has been converted from Brazilian Real (BRL) to their equivalent values in US Dollars (USD) in this plot. We notice that the elasticity of scale is approximately proportional to the logarithm of urban GDP (as indicated by the dashed lines) with the constant of proportionality ($\kappa$) larger for Brazil than the US ($\kappa=0.124\pm0.001$ and $\kappa=0.099\pm0.001$, respectively). We also observe that the elasticity of scale of urban areas with similar GDP is always larger for Brazilian municipalities than US counties.}
\label{fig:fig3}
\end{figure*}

Intriguingly, the translog model predicts that urban areas transit from decreasing to increasing returns to scale of GDP depending on the values of $N$ and $S$. By analyzing the inequality $\varepsilon>1$, we find that whether the product between population and incoming commuters $\Omega = N S$ exceeds or not the critical value $\Omega^* = 10^{(1-\beta_N - \beta_S)/\beta_C} $ determines if an urban area shows decreasing ($\Omega < \Omega^*$) or increasing ($\Omega > \Omega^*$) returns to scale~\cite{ribeiro2019effects}. Moreover, the line defined by $\log S = \log \Omega^* + \log N$ separates two regions in a plane of $\log S$ versus $\log N$, such that urban areas with increasing returns to scale are above this line, and those with decreasing returns to scale are below this line. Figure~\ref{fig:fig3}{a} shows this plane in which the color-coded background refers to the elasticity of scale for each point in the plane and the gray circles stand for values of $\log S$ versus $\log N$ for all US counties and Brazilian municipalities. 

We note that incoming commuters and population are approximately power-law related, that is, $S \sim N^\gamma$ (or $\log S \sim \gamma \log N$) with the exponent $\gamma = 0.87 \pm 0.01$ for the US and $\gamma = 1.06 \pm 0.01$ for Brazil. However, the spread around this power-law trend allows the elasticity of scale to effectively depend on both population and incoming commuters. Thus, urban areas with similar populations (as illustrated for the US urban areas Bethel-AK and Montour-PA, and the Brazilian municipalities Marzagão-GO and Monções-SP) can have significantly different elasticity of scale. Analogously, urban areas with similar number of incoming commuters (as illustrated for the US urban areas McMullen-AK and Hawaii-HI, and the Brazilian municipalities Serra da Saudade-MG and Sant'Ana do Livramento-RS) can also display significantly different elasticity of scale. It is also worth mentioning that the value of $\gamma>1$ observed for Brazil agrees with the idea that the developing stage of the Brazilian transportation infrastructure disproportionately affects small urban areas, creating a more uneven distribution of incoming commuters among Brazilian municipalities. In contrast, the $\gamma<1$ observed for the US indicates a more balanced distribution of incoming commuters and that even small urban areas can have a significant number of incoming commuters.

Using the expression for $\varepsilon$ and the adjusted parameters, we calculate the elasticity of scale for every county and municipality in our data set. Figure~\ref{fig:fig3}{b} depicts the probability distribution of this quantity for the US and Brazil. While both countries have a similar number of urban areas with $\varepsilon>0$ (approximately $63\%$ of US counties and Brazilian municipalities), the probability distribution of $\varepsilon$ for Brazilian municipalities is considerably more skewed to the right. Because of this, the average elasticity of scale of Brazilian urban areas ($\langle \varepsilon \rangle=1.04\pm0.001$) is slightly larger than those in the US ($\langle \varepsilon \rangle=1.02\pm0.001$), and the number of Brazilian urban areas with more extreme values of $\varepsilon$ is significantly larger than in the US (for instance, Brazil has 262 municipalities with $\varepsilon>1.2$ while only 34 US counties meet this criterion). 

Despite the differences between the definitions of usual urban scaling exponent $\beta$ and elasticity of scale $\epsilon$, both quantities measure the response of GDP under changes in population ($\beta$) and in population and incoming commuters ($\epsilon$), such that a qualitative comparison between these quantities is possible. Having made this consideration, the fact that the average elasticity of scale of Brazil is larger than the US one agrees with the negative association between the usual urban scaling exponent and GDP of countries reported in Ref.~\cite{ribeiro2021association} (see Fig.~9 in their supporting information). However, we observe a positive association between elasticity of scale and urban GDP for urban areas in Brazil and US. Figure~\ref{fig:fig3}{c} shows that the elasticity of scale is approximately proportional to the logarithm of the urban GDP ($\varepsilon \sim \kappa \log Y$, with $\kappa$ being a constant of proportionality) for both countries but with Brazil having a slightly sharper association ($\kappa=0.124\pm0.001$ for Brazil and $\kappa=0.099\pm0.001$ for US). It is worth observing that this association can be explained by the relations $S \sim N^\gamma$ and $Y \sim N^\beta$. By solving these latter relations for $N$ and $S$ and plugging the results into the definition of $\varepsilon$, we find $\varepsilon \sim (\beta_C/\beta)(1+\gamma) \log Y$ which in turn implies $\kappa = (\beta_C/\beta)(1+\gamma)$. The estimates of $\kappa$ using this expression ($\kappa \approx 0.09$ for the US and $\kappa \approx 0.16$ for Brazil) are in reasonably good agreement with the values obtained by fitting the relationship shown Fig.~\ref{fig:fig3}{c} with deviations explained by the estimation errors of the other parameters ($\beta$, $\beta_C$, and $\gamma$) and by the systematic underestimation of $S$ by the adjusted relation with $N$ ($S \sim N^\gamma$) in large Brazilian municipalities.

While appearing contradictory at first glance, the differences in the association between $\beta$ and the GDP of countries and the association between $\varepsilon$ and the GDP of urban areas are likely to represent an aggregation effect. As shown in Fig.~\ref{fig:fig3}{c}, Brazilian municipalities with GDP values similar to US counties always have large values of elasticity of scale, which in turn yields an average elasticity of scale larger for Brazil than US. Thus, we expect the elasticity of scale to be associated with the development degree of the urban area as also discussed by Strano \textit{et al.}~\cite{strano2016rich} in the context of usual urban scaling for Eastern and Western European cities. 

\section*{Discussion}

We have investigated the effects of commuting networks on the GDP of urban areas using data from the US and Brazil. Inspired by the economic theory of production functions, we have modeled urban GDP as the output of a production model where population and incoming commuters (one of the simplest features of these networks) are the input variables. These generalized models assume that cities generate wealth by processes that combine their own population and the population of workers attracted from neighboring cities. We have demonstrated that these models significantly refine the description of urban GDP compared with urban scaling theory. This result allowed us to conclude that the deviations observed in the urban scaling are not random variations; instead, we have found that incoming commuters contribute to these deviations and ultimately to the GDP of urban areas.

Beyond yielding more accurate predictions for the GDP of urban areas in the US and Brazil, our modeling approach indicates that population and incoming commuters have an interconnected role in urban wealth production. In contrast to usual urban scaling, our results indicate that changes in urban GDP associated with proportionate changes in population are not independent of the initial population size. Instead, our approach suggests that the GDP of cities with different population sizes and incoming commuters respond differently to proportionate changes in these quantities. Intriguingly, increasing returns to scale of urban GDP is only possible when the product between population and number of incoming commuters exceeds a well-defined threshold. We have found that approximately 63\% of all urban areas in the US and Brazil satisfy the criterion for having increasing returns to scale of GDP. For all remaining 37\% urban areas of both countries, doubling population and number of incoming commuters implies less than doubling their urban GDP. Furthermore, we observe that the GDP of cities with similar population sizes or similar incoming commuters may respond quite differently to proportionate changes in population and incoming commuters. 

When comparing the results between the US and Brazil, we have found that the number of urban areas with high elasticity of scale of GDP is significantly greater in Brazil, making the average elasticity of Brazilian urban areas larger than those in the US. Thus, while Brazilian cities are almost always poorer than the US ones, they usually respond more sharply to proportionate changes in population and incoming commuters. This result agrees with previous findings on urban scaling, which have shown that the GDP of urban areas in rich countries~\cite{ribeiro2021association} or groups of rich cities~\cite{strano2016rich} usually responds less sharply to population changes than their poorer counterparts. However, our results indicate that the elasticity of scale increases with the GDP of Brazilian municipalities and US counties. Thus, the negative association observed between the usual urban scaling exponent and the GDP of countries is likely to represent an aggregation effect of the elasticity of scale of individual urban areas. Moreover, we have found that Brazilian urban areas with GDP comparable to the US ones have significantly larger elasticity of scale. Large values of elasticity of scale are thus more likely to be associated with a lower degree of development than the overall wealth of urban areas.

Finally, we remark that the patterns observed in our study emerge from the interactions between population and number of commuters and by putting forward the idea that cities are not isolated entities. The significantly improved description of urban GDP in terms of population and incoming commuters corroborates this hypothesis and invites more complex approaches to describe urban indicators not only in terms of intra-city processes but through a theory that accounts for inter-city processes and interactions among cities. Other features of these commuting networks could be eventually helpful in describing the GDP or other urban indicators, and phenomenological-like approaches as the ones used here could also be used to identify variables related to city interactions. However, commuting networks are only one facet of city interactions and many other aspects such as the flow of goods, capital, and knowledge among cities are also likely to affect the economic performance of cities. Future investigations can also account for other city indicators affecting urban GDP (such as number of unemployed, education and infrastructure metrics) by generalizing our models to include new variables as well as to understand their interplay. Moreover, recent works have demonstrated that urban scaling is a time-varying phenomenon~\cite{meirelles2018evolution, bettencourt2020interpretation, xu2020scaling} as well as the interaction networks among cities. Despite the limitations in finding dynamic data about commuting networks, it would be interesting to extend our approach to account for dynamical features of city interactions or to use more complex artificial intelligence methods capable of learning and extracting features from commuting networks~\cite{liu2020learning, spadon2020pay}. It is also worth remarking that all results presented here are based on administrative definitions of city (counties for US and municipalities for Brazil, and although obtaining commuting or other network-like data for several definitions of city is challenging, it would also be interesting to investigate how the exponents associated with the Cobb-Douglas and translog models are affected by these different city definitions. Regardless of the city definition and even if we manage to find ideal city boundaries, fact is that cities are not isolated entities. Thus, we hope our work stimulates further research on a theory of cities as complex interconnected systems that is far from equilibrium and depends not only on intra-city networks but also on the environment and connections with the outside world. 

\section*{Methods}

\subsection*{Data sets}

\textit{Commuting networks.} These data sets comprise the number of workers traveling from their residential urban area $v_i$ (a US county or a Brazilian municipality) to another workplace urban area $v_j$. The commuting network $G(V, W)$ is thus the set of urban areas $V$ and directed weighted edges $W$ among them. Each urban area $v_i \in V$ is a node in this network representation and the origin-destination flow is represented by a directed and weighted link $w_{ij} \in W$ starting from the residence location ($v_i$) to the workplace location ($v_j$). The edge weight $w_{ij}$ represents the sum of all commuters moving daily from residence $v_i$ to workplace $v_j$ location. The US data set includes information about the commuting flow among the 3,220 US counties in 2015. We have obtained this data set from the US Census Bureau (USCB)~\cite{uscb2015} which in turn are produced by the American Community Survey (5-Year ACS, 2011-2015). The data set from Brazil includes information about the commuting flow between the 5,565 Brazilian municipalities in the year 2010, and we have obtained it from the Brazilian Institute of Geography and Statistics (IBGE)~\cite{ibge2010micro}.

\textit{Population and GDP data.} We have also collected the population and the urban gross domestic product (GDP) for all US counties from the National Historical Geographic Information System (NHGIS, 5-Year ACS, 2011-2015)~\cite{manson2017ipums}. For the Brazilian municipalities, we have obtained the analogous information for the year 2010 from the Department of Data Processing of the Brazilian Public Healthcare System (DATASUS)~\cite{datasus2010} which are in turn produced by the Brazilian Institute of Geography and Statistics (IBGE).

\subsection*{Ridge regression approach}

While ordinary least-squares is the most common approach for estimating parameters of linear models, this method assumes predictor variables ($\log N$, $\log S$, and $\log N \log S$ in our case) to be uncorrelated. If this assumption does not hold (a problem known as multicollinearity), ordinary least-squares estimates of linear coefficients can be unstable against minor changes in data and display large standard errors. Figure~\ref{fig:fig3}a shows that our predictor variables are correlated and to account for this multicollinearity, we have fitted the Cobb-Douglas (Eq.~\ref{eq:cobb_douglas_scaling}) and translog (Eq.~\ref{eq:translog_scaling}) models to our data using the ridge regression method~\cite{friedman2001elements}. This approach accounts for correlations among the predictor variables by finding the optimal linear coefficients ($\beta_N$, $\beta_S$, and $\beta_c$) that minimize the residual sum of squares plus a penalty term proportional to the sum of the squares of the linear coefficients, where $\lambda$ is the constant of proportionality. We have estimated the optimal value of $\lambda=\lambda^*$ by minimizing the mean squared error with the leave-one-out cross-validation method. For this procedure, we have first standardized all predictors variables before finding $\lambda^*$ to ensure that the regularization term is uniformly applied to all predictors. This regularization term reduces the variance and provides more reliable and stable estimates of the model parameters~\cite{friedman2001elements}. We have used the implementation of the ridge regression method available in the Python module \texttt{scikit-learn}~\cite{pedregosa2011scikitlearn}.

\bibliography{commuters}

\clearpage

\setcounter{page}{1}
\setcounter{figure}{0}
\rfoot{\small\sffamily\bfseries\thepage/1}%
\makeatletter 
\renewcommand{\thefigure}{S\@arabic\c@figure}
\renewcommand{\thetable}{S\@arabic\c@table}

\begin{center}
\large{Supplementary Information for}\\
\vskip1pc
\large{\bf Commuting network effect on urban wealth scaling}\\
\vskip1pc
\normalsize{Luiz G. A. Alves, Diego Rybski, and Haroldo V. Ribeiro}\\
\vskip1pc
\normalsize{Scientific Reports, 2021}\\
\end{center}
\vskip2pc

\begin{figure}[!h]
\centering
\includegraphics[width=0.8\linewidth]{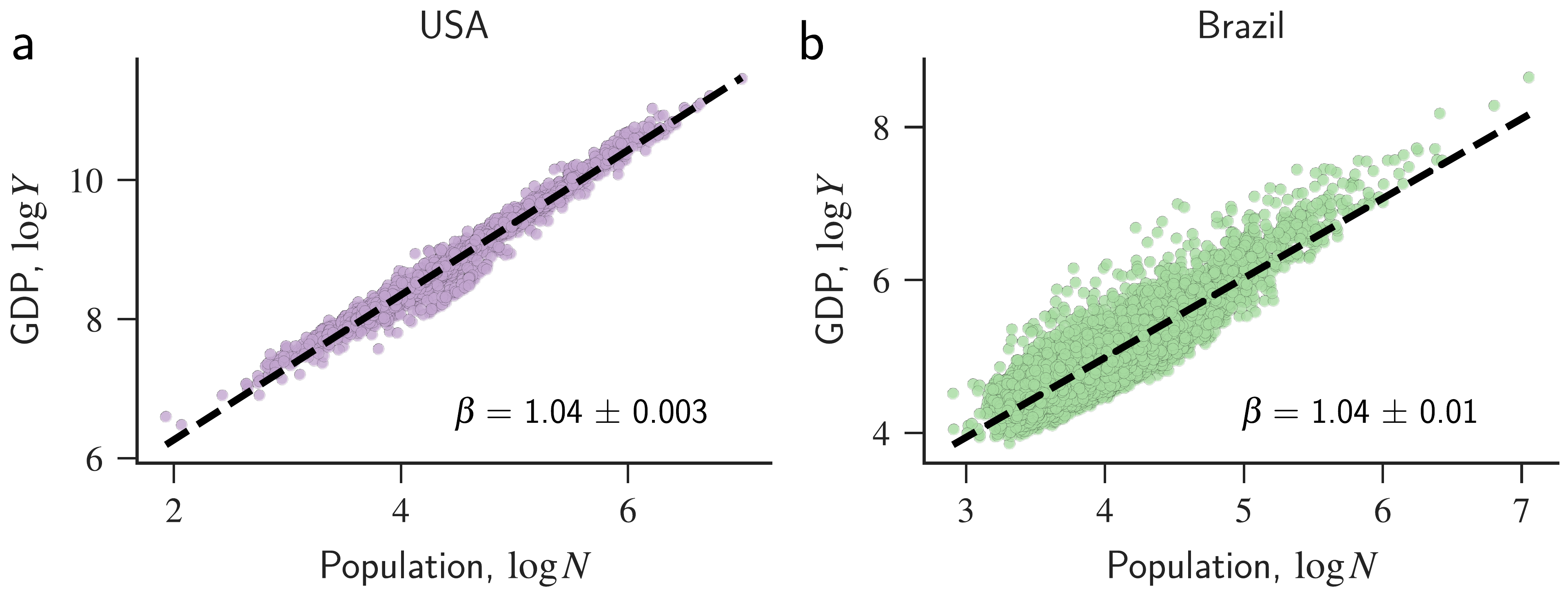}
\caption{Scaling relationships between urban GDP ($Y$) and population size ($N$) for US counties (purple markers of panel a) and  Brazilian municipalities (green markers of panel b). The dashes lines represent power-law fits (Eq. 1) with exponents $\beta$ indicated within each panel.}
\label{sfig:1}
\end{figure}

\end{document}